\documentclass[useAMS,usenatbib]{mn2e}

\usepackage{amsmath,array,amsfonts}
\usepackage{amssymb}
\usepackage{graphicx}
\usepackage{natbib}
\usepackage{pstricks}
\usepackage{verbatim}
\usepackage{multirow}
\usepackage{aas_macros}
\usepackage{bbold}

\newcommand{\ltsima} {$\; \buildrel < \over \sim \;$}
\newcommand{\gtsima} {$\; \buildrel > \over \sim \;$}
\newcommand{\lta} {\lower.5ex\hbox{\ltsima}}
\newcommand{\gta} {\lower.5ex\hbox{\gtsima}}

\title[Using CMB polarization to constrain the anomalous nature of the Cold Spot]{Using CMB polarization to constrain the anomalous nature of the Cold Spot with an incomplete sky-coverage}

\author[R. Fern\'{a}ndez-Cobos et al.]{R. Fern\'{a}ndez-Cobos$^1$$^,$$^2$\thanks{e-mail:cobos@ifca.unican.es}, P. Vielva$^1$, E. Mart\'inez-Gonz\'alez$^1$, M. Tucci$^3$, M. Cruz$^4$\\
$^1$     Instituto de F\'isica de Cantabria, CSIC-Universidad de Cantabria, Avda. de los Castros s/n, 39005 Santander, Spain.\\
$^2$     Dpto. de F\'isica Moderna, Universidad de Cantabria, Avda. los Castros s/n, 39005 Santander, Spain.\\
$^3$     D\'epartement de Physique Th\'eorique, Universit\'e de Gen\`eve, 24, Quai E. Ansermet, 1211 Gen\`eve 4, Switzerland.\\
$^4$     Dpto. de Matem\'aticas, Estad\'istica y Computaci\'on, Universidad de Cantabria, Avda. los Castros, s/n, 39005 Santander, Spain.}

\date{Accepted  Received ; in original form }

\begin{document}

\maketitle

\begin{abstract}
Recent results of the ESA \textit{Planck} satellite have confirmed the existence of some anomalies in the statistical distribution of the cosmic microwave background (CMB) anisotropies. One of the most intriguing anomalies is the Cold Spot, firstly detected in the \textit{WMAP} data by \citet{Vielva2004}. In a later paper, \citet{Vielva2011} developed a method to probe the anomalous nature of the Cold Spot by using the cross-correlation of temperature and polarization of the CMB fluctuations. Whereas this work was built under the assumption of analysing full-sky data, in the present paper we extend such approach to deal with realistic data sets with a partial sky-coverage. In particular, we exploit the radial and tangential polarization patterns around temperature spots. We explore the capacity of the method to distinguish between a standard Gaussian CMB scenario and an alternative one, in which the Cold Spot arises from a physical process that does not present correlated polarization features (e.g., topological defects), as a function of the instrumental-noise level. Moreover, we consider more in detail the case of an ideal noise-free experiment and those ones with the expected instrumental-noise levels in \textit{QUIJOTE} and \textit{Planck} experiments. We also present an application to the 9-year \textit{WMAP} data, without being able to obtain firm conclusions, with a significance level of $32\%$. In the ideal case, the alternative scenario could be rejected at a significance level of around $1\%$, whereas for expected noise levels of \textit{QUIJOTE} and \textit{Planck} experiments the corresponding significance levels are $1.5\%$ and $7.4\%$, respectively.    
\end{abstract}
\section{Introduction}
\label{sec:Introduction}
Under the current inflationary frame \citep{Starobinsky1980, Guth1981, Linde1982} of the cosmological concordance model ($\Lambda$CDM), the statistical properties of the cosmic microwave background (CMB) anisotropies are a reflection of the features of the primordial density fluctuations. In particular, standard models of inflation predict that these anisotropies are described by an almost Gaussian, homogeneous and isotropic random field. However, some hints of anomalous behaviour regarding this Gaussian pattern have been observed first by \textit{WMAP} \citep{Spergel2003, Schwarz2004, Vielva2004, Land2005, Rossmanith2009} and, more recently, by \textit{Planck} \citep{PlanckXXIII2013}. These findings become crucial in order to discard alternative proposals, since non-standard inflation scenarios \citep{Linde1997, Bartolo2004, Bernardeau2002, Gupta2002, Gangui2002, Acquaviva2003} and cosmological defect models \citep{Turok1990, Durrer1999} usually predict non-Gaussian fields. One of the most relevant topics in the context of these anomalies was the detection of a non-Gaussian cold spot in the southern hemisphere ($l=209^{\circ}$, $b=57^{\circ}$) using a wavelet analysis of the \textit{WMAP} data \citep{Vielva2004, Cruz2005}. Its existence was confirmed by several authors \citep{Mukherjee2004, Cayon2005, McEwen2005, Rath2007, Vielva2007, Pietrobon2008, Gurzadyan2009, Rossmanith2009}  through different techniques, and, recently, in \textit{Planck} data \citep{PlanckXXIII2013}.

Many theoretical explanations have been proposed to justify the presence of the Cold Spot (CS), namely second order gravitational effects \citep{Tomita2005, Tomita2008}, contamination from foreground residuals \citep{Liu2005, Coles2005}, a finite universe model \citep{Adler2006}, large voids \citep{Inoue2006, Rudnick2007, Granett2008, Garcia2008, Masina2009}, the collision of cosmological bubbles \citep{Chang2009}, textures in a brane-world model \citep{Cembranos2008} or a non-Gaussian modulation \citep{Naselsky2010}. Nevertheless, several works have shown that some of these explanations are very unsatisfactory \citep{Cruz2006, Smith2010} because many of the previous arguments require very especial conditions, such as a peculiar orientation of large voids or a particular proportion of foreground residuals. 

An alternative hypothesis was presented in \citet{Cruz2007} which maintains that the CS could be caused by the non-linear evolution of the gravitational potencial created by a cosmic texture. This work shows a comparison between the texture hypothesis and the standard Gaussian frame, concluding by a Bayesian analysis that the first one is preferred.

Indeed, cosmic textures are theoretically well-motivated, although their contribution to the CMB anisotropies has been proven to be subdominant \citep{Bevis2004, Urrestilla2008}. 

Although promising, further tests are needed to prove the existence of cosmic textures. As the texture model predicts an expected number of cosmic textures with an angular scale greater than a certain size $\theta$, one of the next steps in the investigation is the detection of other candidates \citep[e.g.,][]{Gurzadyan2009, Vielva2007, Pietrobon2008}. Furthermore, the pattern of the CMB lensing introduced by a texture is known and it should be measured, if present, by high resolution CMB experiments such as the South Pole Telescope \citep[e.g.,][]{Das2009}. Finally, a lack of polarization signal, in comparison with the levels associated to a Gaussian and isotropic random field, is predicted by the texture scenario. 

Starting from this difference in the polarization signal, \citet{Vielva2011} presented a method to distinguish between both hypotheses: a large fluctuation of a Gaussian and isotropic random field and an anomalous feature caused, for instance, by a cosmic texture. The criterion used to discriminate between both cases was based on the difference between the cross-correlation of the radial profiles in the temperature $\mu_T(\theta)$ and the E-mode polarization $\mu_E(\theta)$. That estimator is grounded in the possibility of having an E-mode map of tens of degrees centred in the position of the CS. However, such map is hard to be obtained from current and incoming polarization data sets, due to the incomplete sky-coverage.

To avoid this problem, we present an extension of the approach, by considering the radial and tangential patterns of the Stokes polarization parameters around the position of the CS. These quantities are more natural, as they relate in a more direct way to the measurements obtained by polarization experiments.

This paper is structured as follows. In section \ref{sec:method} we describe how the cross-correlation is taken into account in our approach. We present the methodology used to discriminate between the Gaussian and texture hypotheses in Section \ref{sec:method}. In Section \ref{sec:results} we explore the scopes of our method with CMB simulations with different noise levels. An application to 9-year \textit{WMAP} data is shown in Section \ref{sec:data}. Finally, conclusions are exposed in Section \ref{sec:Conclusions}.

%
\section{Characterisation of the cross-correlation \textit{TE}}
\label{sec:characterization}
The main problem that arises in this work is to distinguish between two different scenarios that are called null and alternative hypotheses. On the one hand, we denote as null hypothesis ($\mathrm{H_0}$) the case in which all the CMB fluctuations (including the CS) are caused by a standard Gaussian and isotropic random field. On the other hand, the alternative hypothesis ($\mathrm{H_1}$) is associated with the case where the CMB fluctuations are generated by the standard-model mechanisms but there is a non-negligible contribution due to a physical process that does not produce a correlated pattern between temperature and polarization, such as topological defects. 

As shown in \citet{Vielva2011}, the cross-correlation between the temperature and the E-mode of polarization around the location of the CS can be used to discriminate between these two hypotheses. The $\mathrm{H_1}$ scenario would show a lack of correlation between the temperature spot and the associated polarization as compared to $\mathrm{H_0}$. In this case, it is assumed that the CS is generated by a secondary anisotropy of the CMB, modified by a non-linear evolution of a gravitational potencial. This evolution could be due to, for instance, a collapsing cosmic texture.

\begin{figure}
\begin{center}
\includegraphics[scale=0.35]{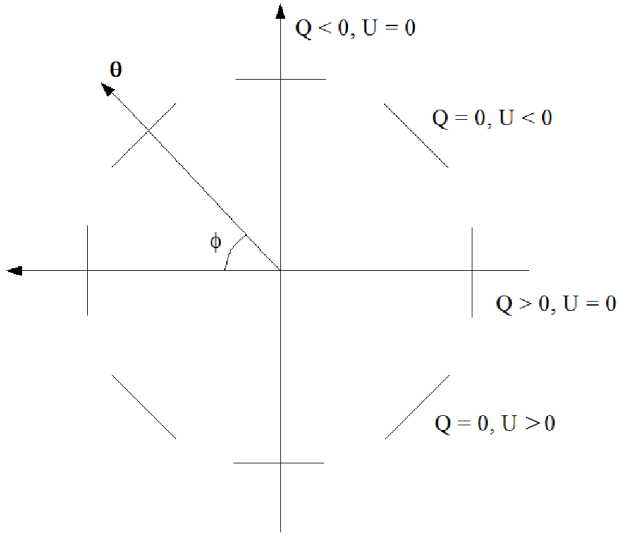}
\end{center}
\caption{Parametrization of the Q and U Stokes parameters rotation.}
\label{fig:rot_param}
\end{figure}

In practice, obtaining a reliable E-mode map is very complicated, since the full-sky information cannot be recovered. Therefore, instead of considering directly the cross-correlation as the product of temperature and polarization radial profiles, as done in \citet{Vielva2011}, we use a locally-defined rotation of the Stokes parameters:
\begin{equation} \begin{array}{ccc}
Q_\mathrm{r}\left(\boldsymbol{\theta}\right) & = & -Q\left(\boldsymbol{\theta}\right)\cos{(2\phi)}-U\left(\boldsymbol{\theta}\right)\sin{(2\phi)}  \\
U_\mathrm{r}\left(\boldsymbol{\theta}\right) & = & Q\left(\boldsymbol{\theta}\right)\sin{(2\phi)}-U\left(\boldsymbol{\theta}\right)\cos{(2\phi)}  \\
\end{array},
\label{eq:def_qr}
\end{equation}
where $\boldsymbol{\theta} = \theta \left( \cos{\phi}, \sin{\phi}\right)$ and $\phi$ is the angle defined by the line that connects the temperature spot at the centre of the reference system and a position at an angular distance $\theta$ from the centre, as is shown in Figure \ref{fig:rot_param}.

The new Stokes parameters are expressed in another coordinate system that is rotated by $\phi$ respect to the $\mathrm{Q}$ and $\mathrm{U}$ frame. This definition was first introduced by \citet{Kamionkowski1997} and it is a way to decompose the polarization signal into a radial and a tangential component in a local frame (note that the Stokes parameter axes live in the tangential plane). The same definition is used by \citet{Komatsu2011} to compare the $\mathrm{Q_r}$ and $\mathrm{U_r}$ patterns around temperature cold and hot spots in the \textit{WMAP} data. 

The previous expressions are a flat-sky approximation, valid in a region of $\sim 5^{\circ}$ of radius from the reference centre \citep{Komatsu2011}. To overcome this difficulty, we transform $\mathrm{Q}$ and $\mathrm{U}$ equivalently by rotating the map such that the temperature spot is located on the north pole. In this particular configuration, the angle $\phi$ coincides with the longitude of the sphere and the polarization axes of $\mathrm{Q}$ are naturally radial or tangencial respect to the origin of the reference system at the north pole. Therefore, we can make the identification $\mathrm{Q} \equiv \mathrm{Q_r}$.

A sky map of $\mathrm{Q_r}$ is always referred to a centre position, so it is meaningless beyond the local interpretation. The $\mathrm{Q_r}$ radial profile $\mu_{\mathrm{Q_r}}$ with respect to a centre position $\mathbf{x}$ is defined as:
\begin{equation}
\mu_{Q_\mathrm{r}}\left(\mathbf{x}, \theta \right) = \dfrac{1}{N_{\theta}} \sum_i{Q_\mathrm{r}\left(\mathbf{x}_i\right)}, 
\end{equation}
where the sum is extended over the positions $\mathbf{x}_i$ which are at an angular distance $[\theta - \frac{\Delta \theta}{2}, \theta + \frac{\Delta \theta}{2}]$ from the centre position $\mathbf{x}$. The total number of positions considered in the equidistant ring of width $\Delta \theta$ is denoted by $N_{\theta}$. 

The stacked $\mathrm{Q_r}$ radial profiles, $\bar{\mu}_{Q_{\mathrm{r}}}$, are related to the cross-power spectrum $C_{\ell}^{\mathrm{TE}}$ by an integral with a kernel $f(\ell, \theta)$ that depends on the angular distance and the multipole index \citep[see][Appendix B, for more details]{Komatsu2011}:
\begin{equation}
\bar{\mu}_{Q_\mathrm{r}} (\theta) = \int{f(\ell, \theta) C_{\ell}^{\mathrm{TE}} d\ell}.
\end{equation}
 
The most important point to justify our approach is this dependence, i.e., that the cross-correlation information between temperature and E-mode polarization is contained in a quantity calculated as the stacked $\mathrm{Q_r}$ radial profiles. 

\begin{figure}
\includegraphics[scale=0.45]{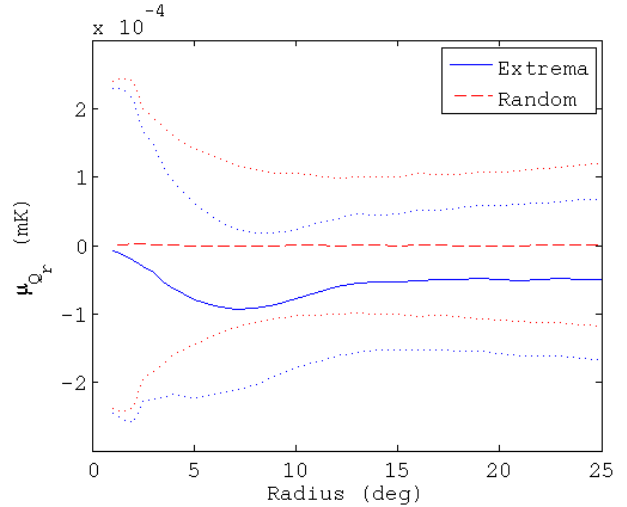}
\caption{Mean $\mathrm{Q_r}$ radial profile at hot extrema positions ($\mathrm{H_0}$), with an amplitude in the temperature maps, at least, as large as the one of the CS (solid blue line), and at random positions ($\mathrm{H_1}$) in the temperature maps (dashed red line). Their corresponding dispersion is shown by dotted lines.}
\label{fig:profiles}
\end{figure}

We show in Figure \ref{fig:profiles} the mean value and dispersion of the $\mathrm{Q_r}$ radial profiles for two different cases computed with CMB simulations. The first one represents a selection of profiles associated to positions $\mathbf{x}_{\mathrm{ext}}$ where a hot spot, at least as extreme as the CS, is identified in the CMB temperature field. The selection criteria is a threshold over $4.45$ times the dispersion of the Spherical Mexican Hat Wavelet (SMHW) coefficients at a wavelet scale $R = 250$ arcmin in absolute value \citep[see][for details]{Vielva2004}. This limit value is calculated as the amplitude (in absolute value) of the CS in the wavelet coefficient of the combined map $\mathrm{Q+V}$ of the 9-year \textit{WMAP} data degraded to a \textit{HEALPix} resolution of $\mathrm{N_{side}}=64$ \citep{Gorski2005}. The second case corresponds to the $\mathrm{Q_r}$ radial profiles referred to randomly-selected central positions $\mathbf{x}_{\mathrm{rnd}}$ in the CMB field. The highest discriminatory-power regime seems to be about a scale of $\theta \sim 7^{\circ}$. 

\begin{figure*}
\includegraphics[scale=0.17]{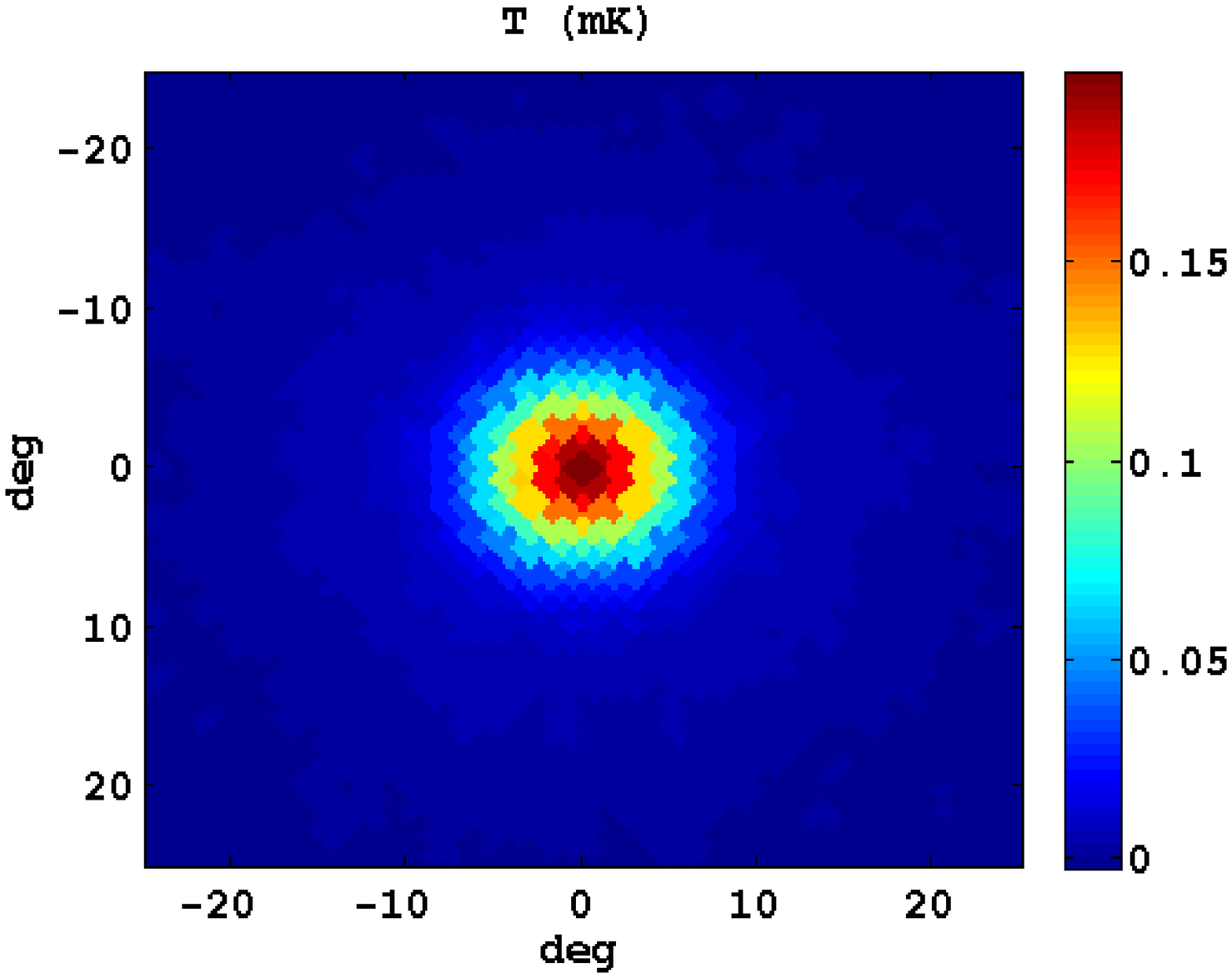}
\includegraphics[scale=0.17]{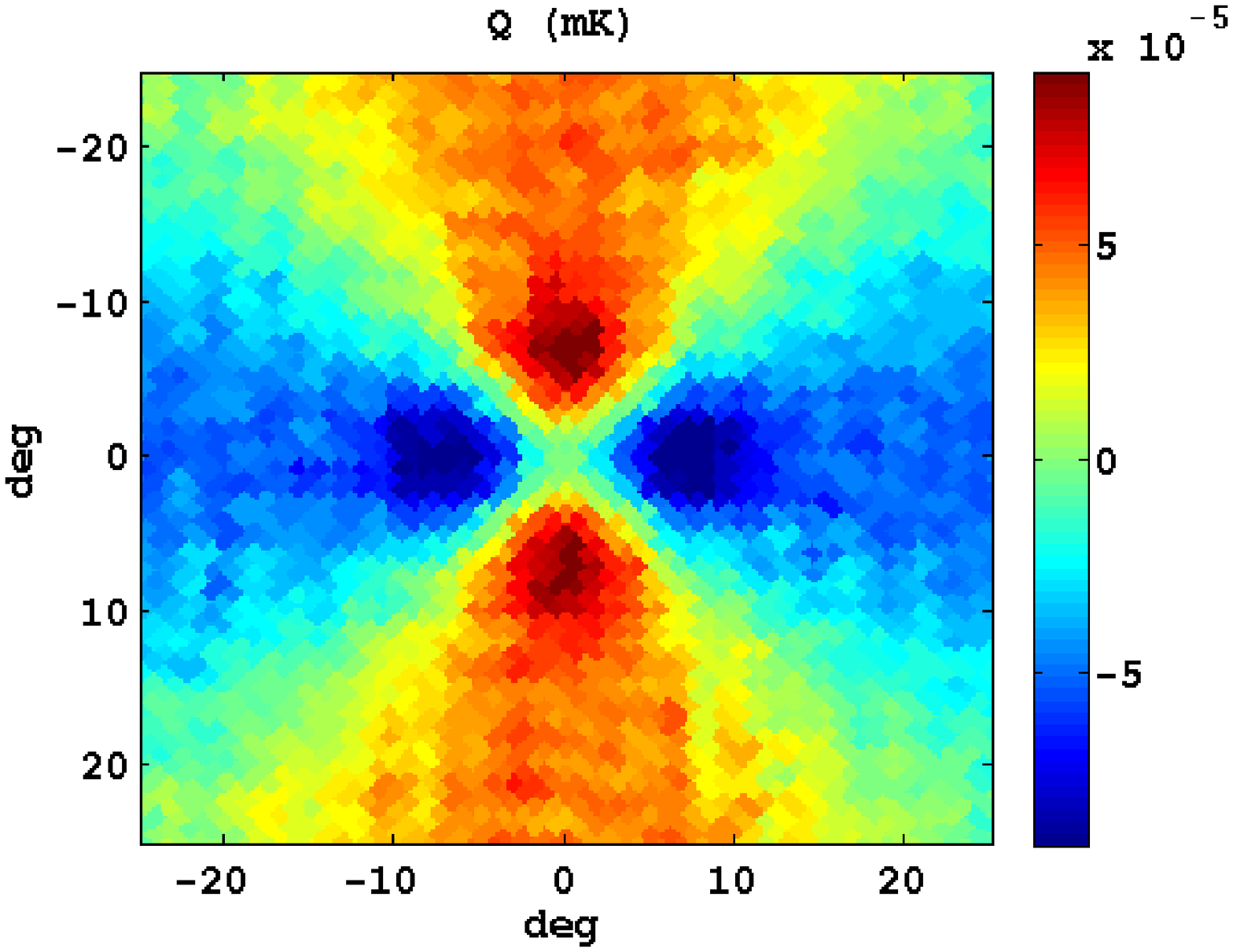}
\includegraphics[scale=0.17]{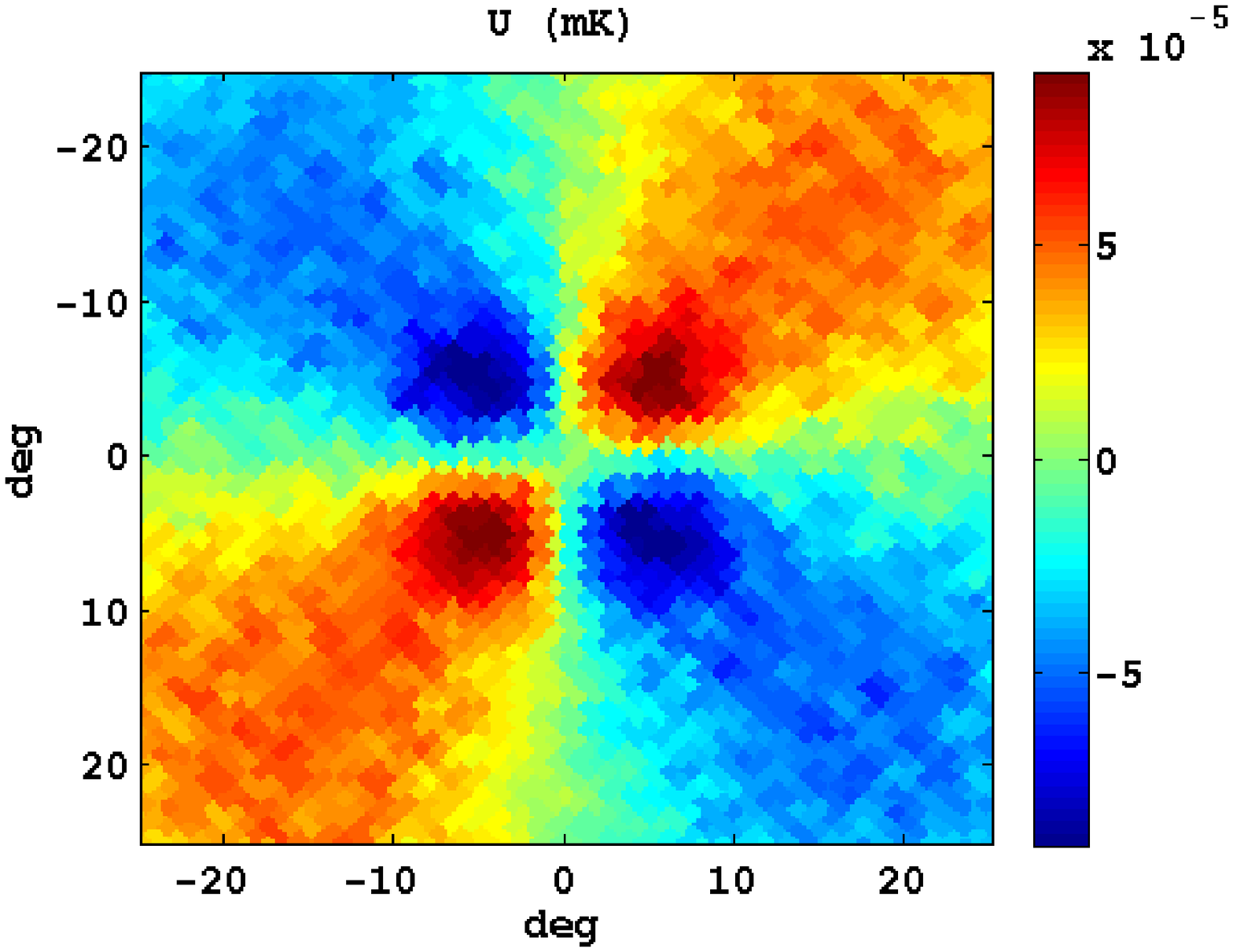}
\includegraphics[scale=0.17]{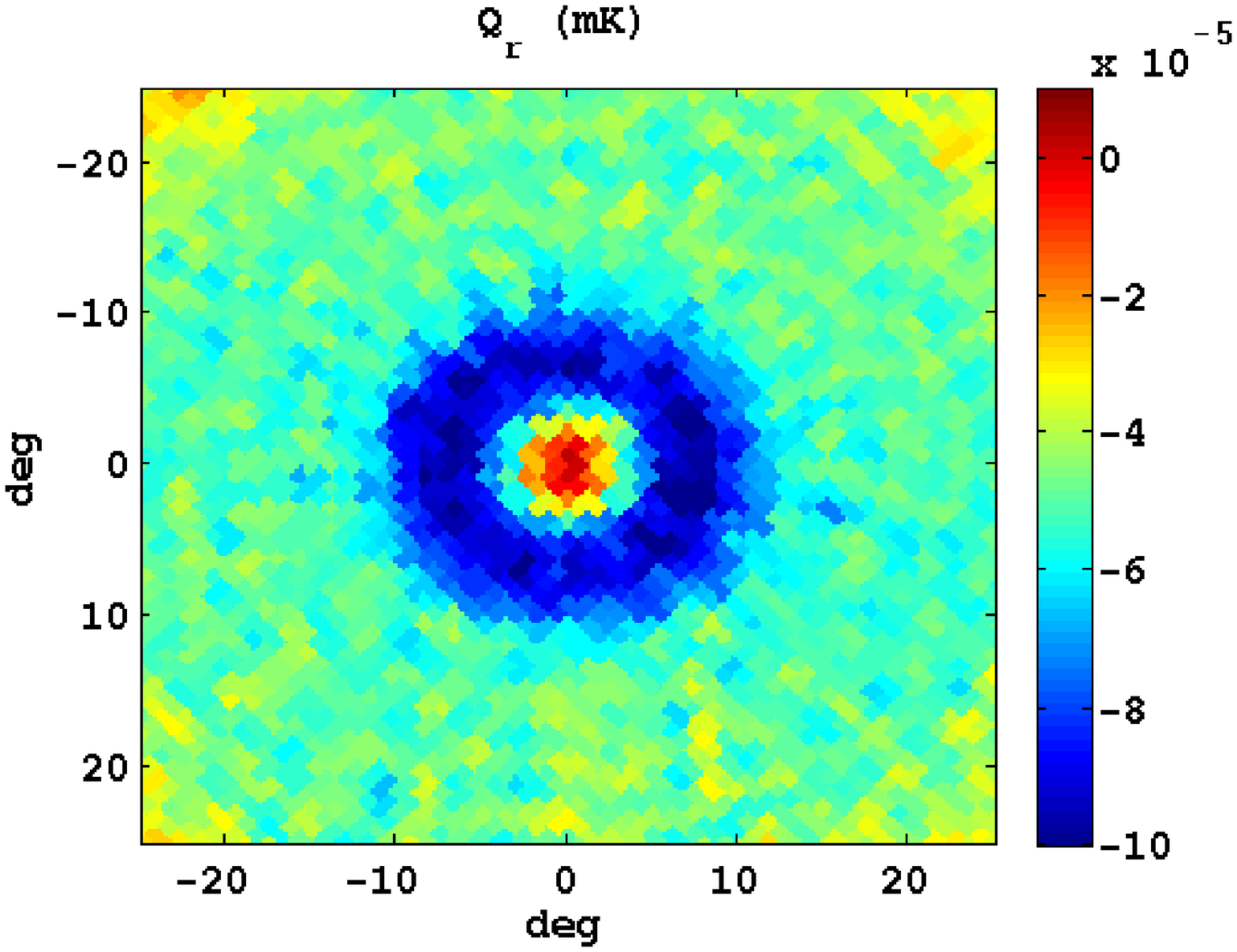}
\includegraphics[scale=0.17]{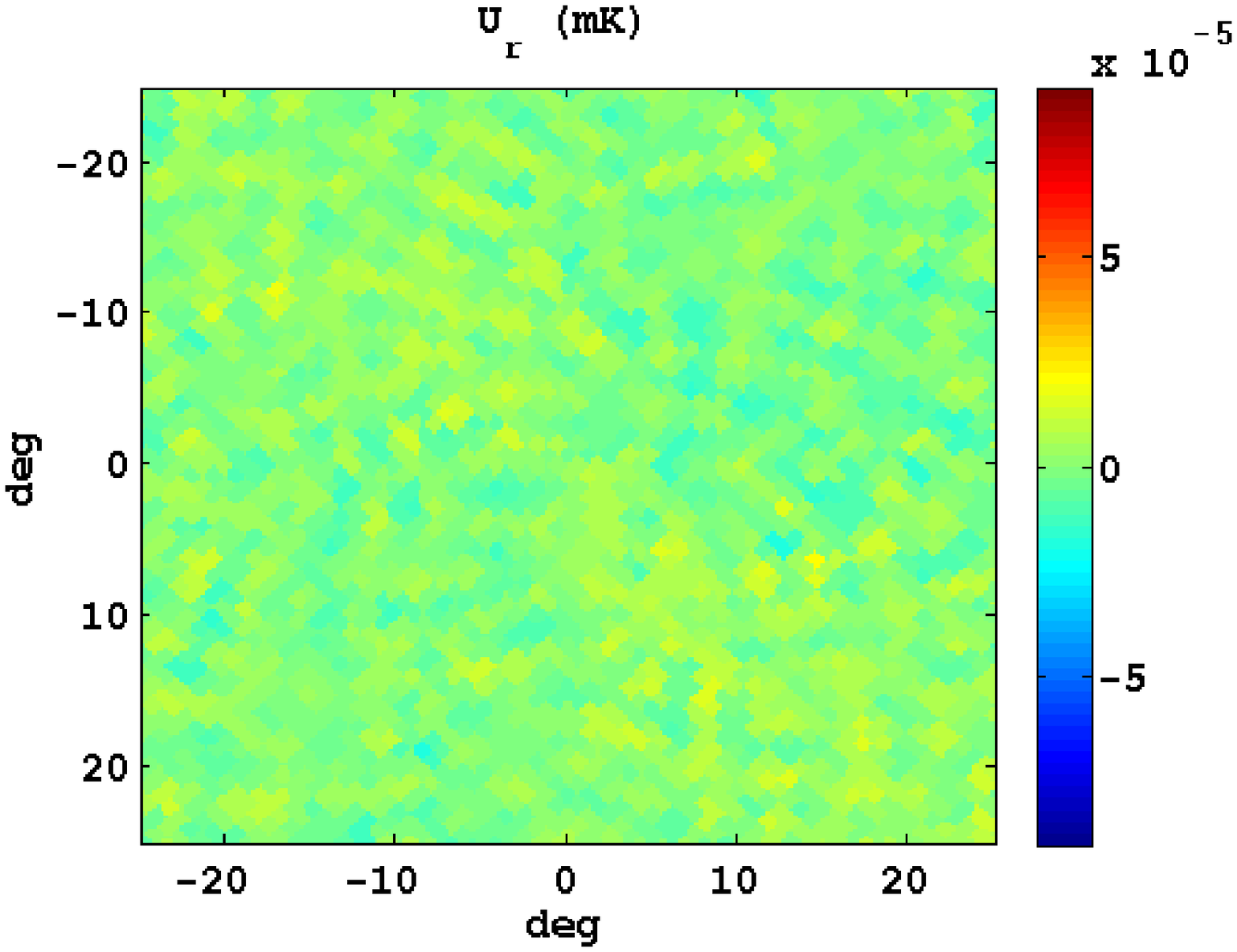}
\includegraphics[scale=0.17]{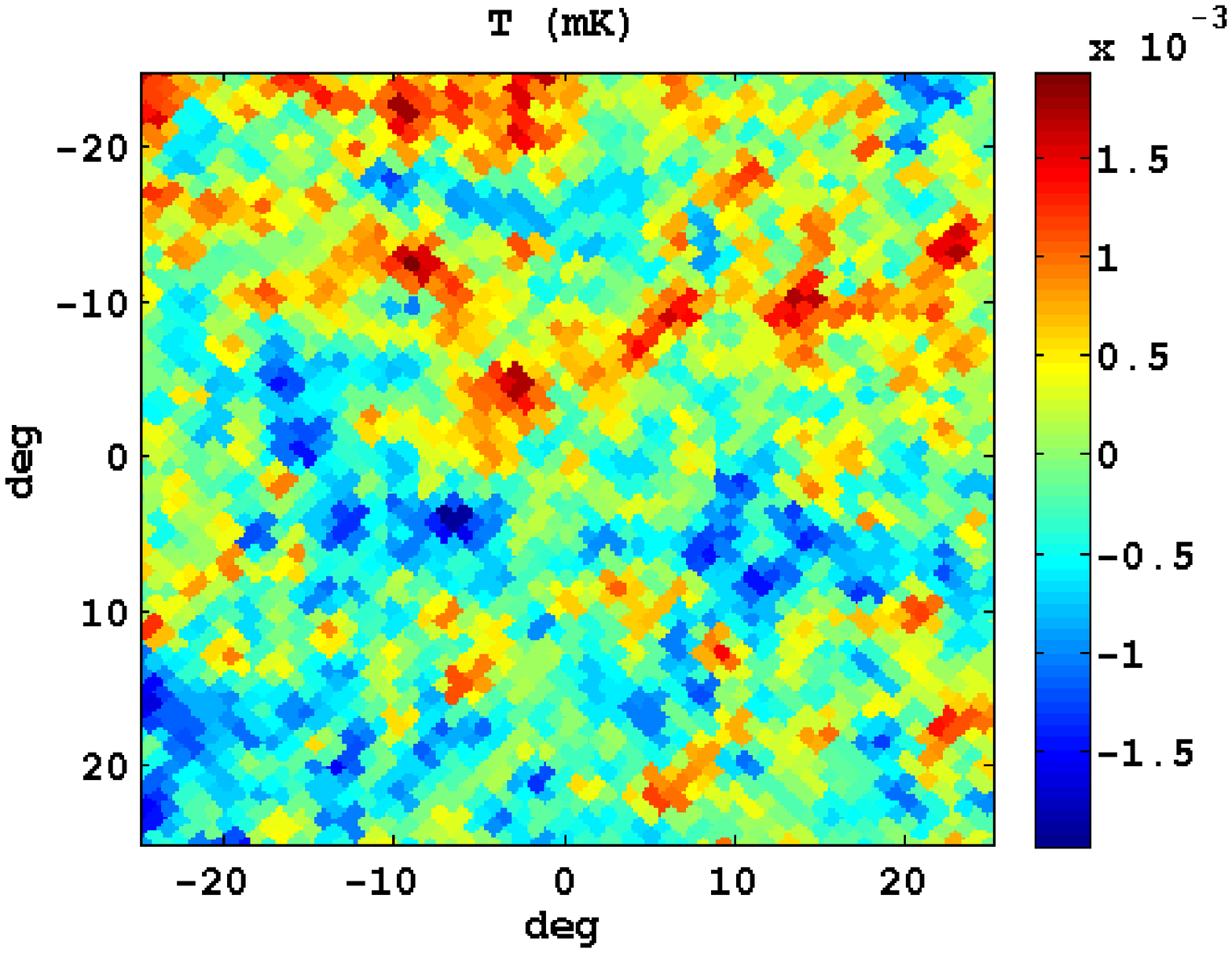}
\includegraphics[scale=0.17]{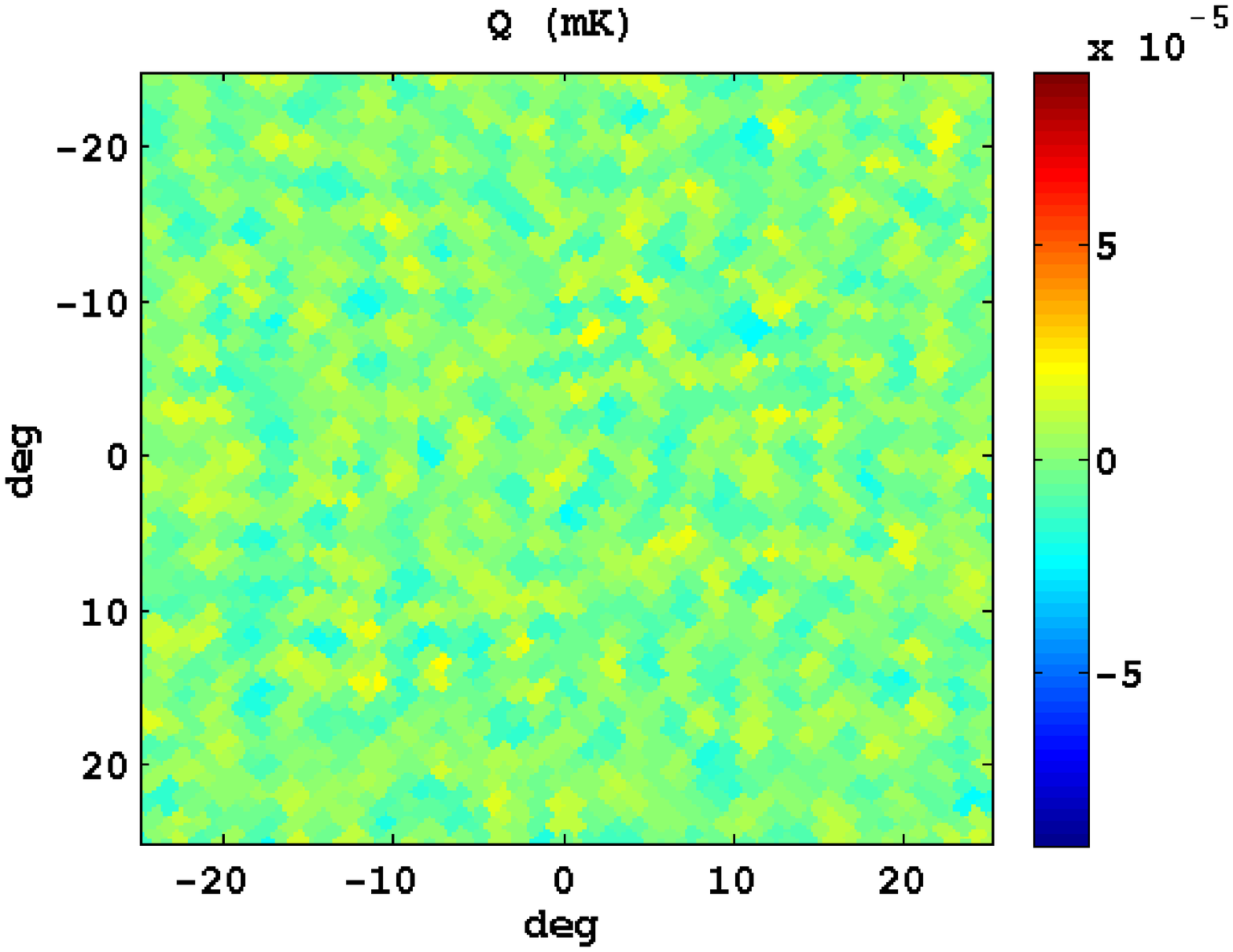}
\includegraphics[scale=0.17]{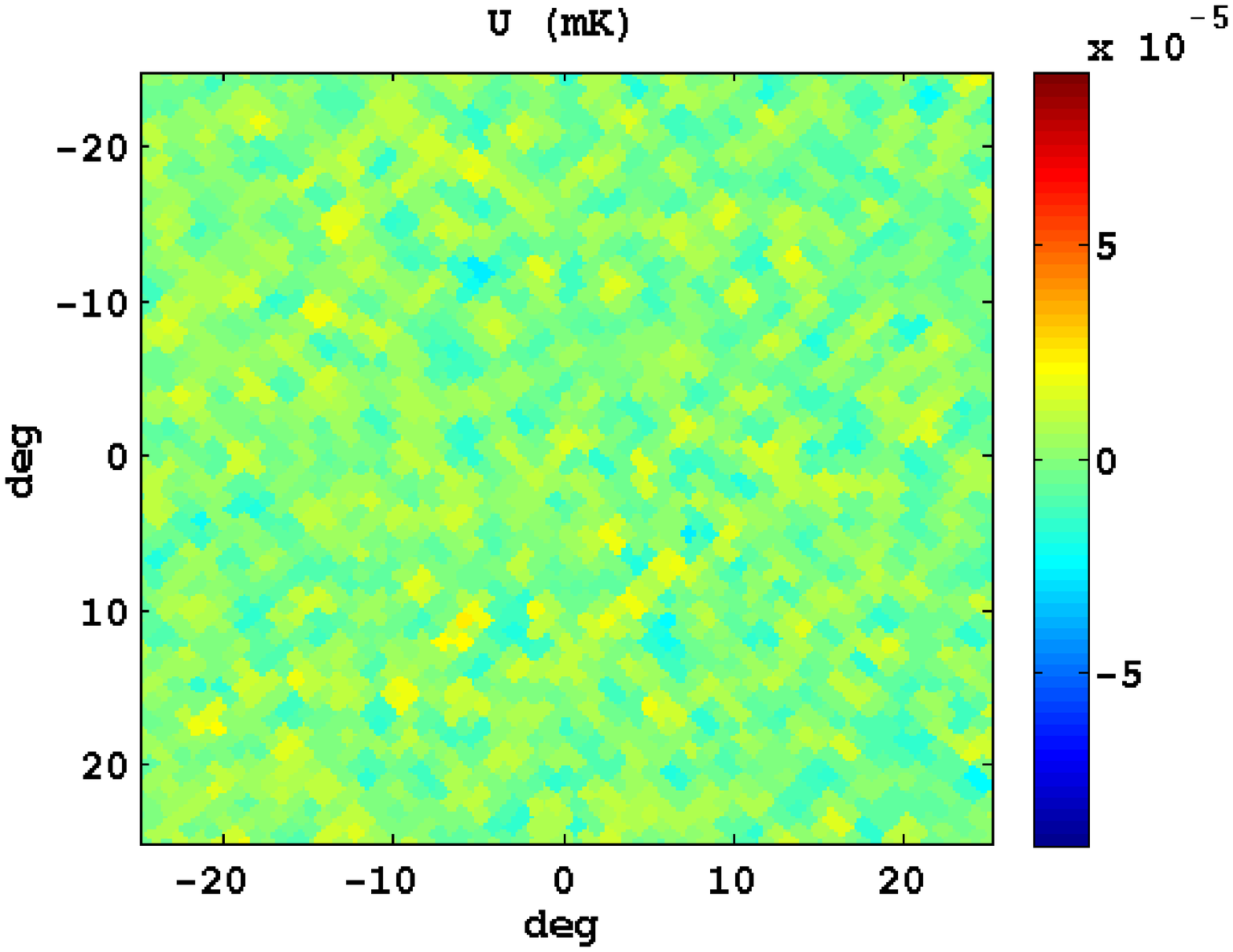}
\includegraphics[scale=0.17]{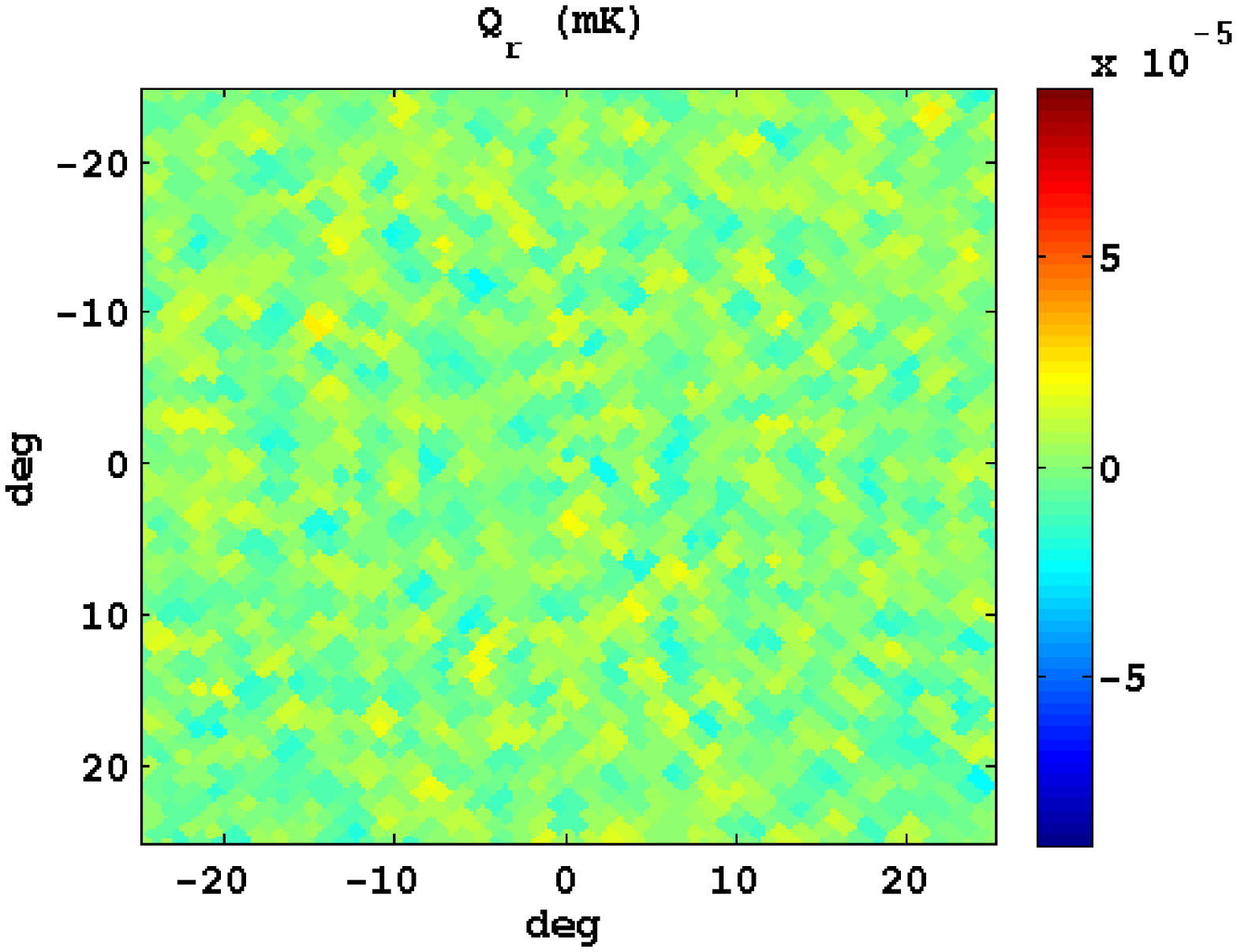}
\includegraphics[scale=0.17]{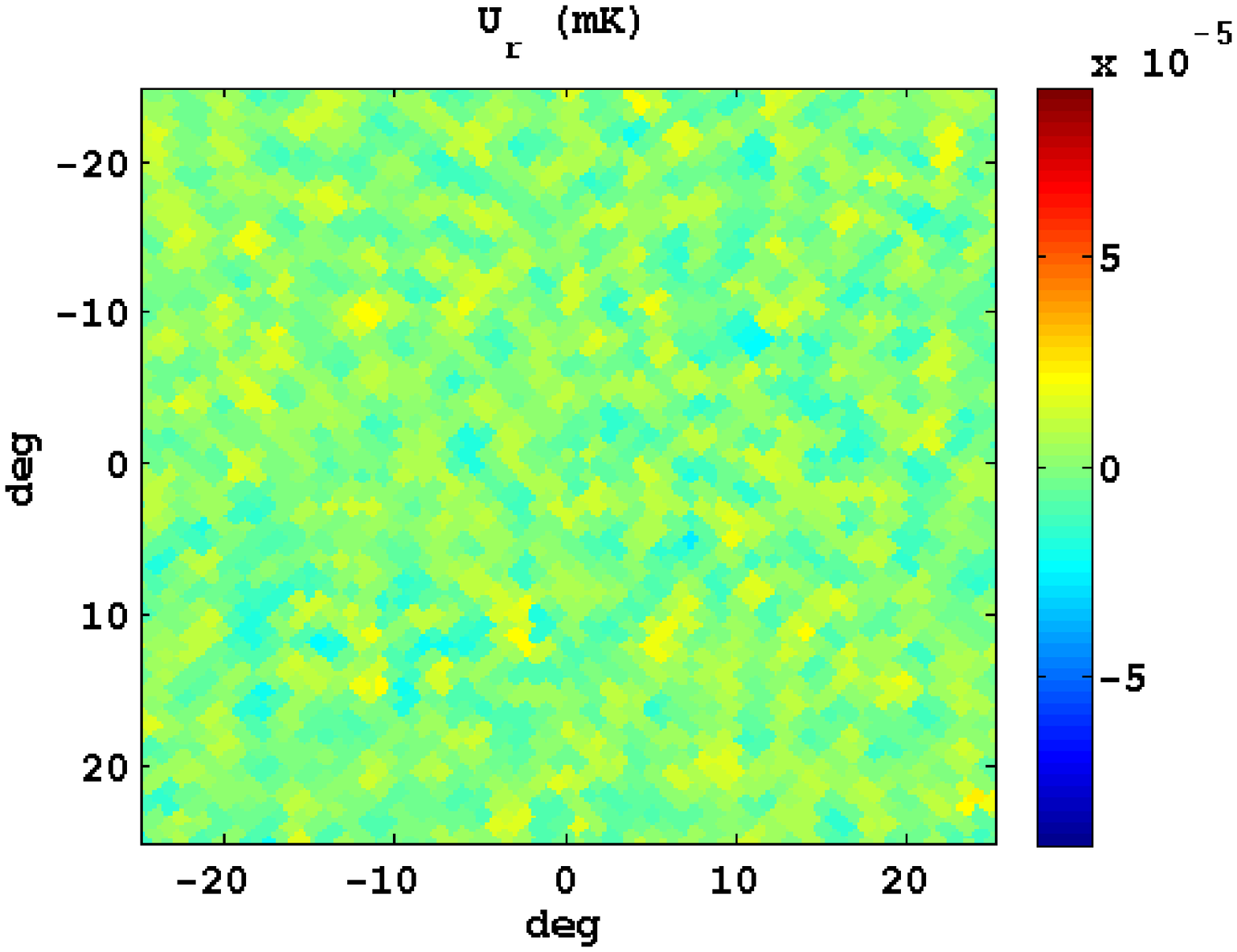}
\caption{From left to right: stacked images of $\mathrm{T}$, $\mathrm{Q}$, $\mathrm{U}$, $\mathrm{Q_r}$ and $\mathrm{U_r}$ Stokes parameters. The upper row corresponds to a stacking of $11000$ simulations in extrema positions (hot spots) and the bottom one contains the analogous panels selecting random positions in temperature.}
\label{fig:stacking}
\end{figure*}

In Figure \ref{fig:stacking}, we show the stacked patterns for the standard Stokes parameters, as well as $\mathrm{Q_r}$ and $\mathrm{U_r}$ for both cases: the upper panels corresponds to hot spot positions as intense in absolute value as the CS and in the bottom ones random locations are plotted. Note that similar but symmetric results would be obtained for cold spots.

The mean radial profiles are obtained by averaging over $11000$ simulations in both cases: using a set of simulations centred in a feature as extreme as the CS, and another one taking random positions. For this study, the first step is to filter the temperature map of the CMB Gaussian simulations with the SMWH at a scale $R = 250$ arcmin. Then, we search a feature as intense as the CS in the wavelet coefficient map. If such trait is not present, we discard this simulation and generate a new one. Nevertheless, if a CS-like feature is found, we compute the $\mathrm{Q_r}$ maps centred in its location $\mathbf{x}_{\mathrm{ext}}$, and calculate the radial profile $\mu_{Q_\mathrm{r}}(\mathbf{x}_{\mathrm{ext}},\theta)$ referred to this certain position. Moreover, the radial profile $\mu_{Q_\mathrm{r}}(\mathbf{x}_{\mathrm{rnd}},\theta)$ is computed taking a random position as reference.

\section{Methodology}
\label{sec:method}
In this section we adapt the formalism used in \citet{Vielva2011} to distinguish between the two hypotheses that we are considering: the standard Gaussian and isotropic option (as null hypothesis, $\mathrm{H_0}$) and a non-standard model proposed as a superposition of a contribution due to a physical mechanism which does not produce correlation between temperature and polarization, such as topological defects, and the CMB fluctuations of the standard model (as alternative hypothesis, $\mathrm{H_1}$).

\subsection{The estimator}
Under the null hypothesis assumption $\mathrm{H_0}$, the cross-correlation pattern between temperature and the polarization E-mode at positions $\mathbf{x}_{\mathrm{ext}}$ (as mentioned, where a CS-like feature is located in the CMB temperature map) is reflected in the $\mathrm{Q_r}$ radial profile $\mu_{Q_\mathrm{r}}$. We can represent the hypothesis with a vector $\xi_{\mathrm{H_0}}$ of $\mathrm{n_r}$ components, where $\mathrm{n_r}$ is the number of rings considered at different distances $\theta_i$ from the centre $\mathbf{x}_{\mathrm{ext}}$:
\begin{equation}
\xi_{\mathrm{H_0}}(i) \equiv \mu_{Q_\mathrm{r}}(\mathbf{x}_{\mathrm{ext}},\theta_i).
\end{equation}
We use values of $\theta$ from $1^{\circ}$ to $5^{\circ}$, separated by $0.5^{\circ}$, and from $5^{\circ}$ to $25^{\circ}$ separated by $1.0^{\circ}$, which represent a total of $29$ rings with $\frac{\Delta \theta}{2}= 0.5^{\circ}$ of width.

Under the assumption of the alternative hypothesis $\mathrm{H_1}$, the CS is not generated by the standard Gaussian and isotropic field but arises due to a secondary anisotropy (e.g., a cosmic texture). In this scenario, there is not any expected correspondence between the temperature extreme and the polarization signal. We can translate this fact considering random field values (mutually consistent) in $\mathrm{Q}$ and $\mathrm{U}$ Stokes parameters. Therefore, the alternative hypothesis can be expressed as:
\begin{equation}
\xi_{\mathrm{H_1}}(i) \equiv \mu_{Q_\mathrm{r}}(\mathbf{x}_{\mathrm{rnd}},\theta_i).
\end{equation}

In both cases ($\gamma=0,1$), we can compute the mean value of these radial profiles $\bar{\xi}_{\mathrm{H_{\gamma}}}$ and a covariance matrix $C_{\mathrm{H_{\gamma}}}$ as:
\begin{equation}
\bar{\xi}_{\mathrm{H_{\gamma}}}(i)  =  \dfrac{1}{N_s} \sum_{n=1}^{N_s}{\xi_{\mathrm{H_{\gamma}},n}(i)} 
\end{equation}
\begin{equation}
C_{\mathrm{H_{\gamma}}}(j,k)  =  \dfrac{1}{N_s} \sum_{n=1}^{N_s}{\left[ \xi_{\mathrm{H_{\gamma}}}(j) - \bar{\xi}_{\mathrm{H_{\gamma}}}(j) \right]  \left[ \xi_{\mathrm{H_{\gamma}}}(k) - \bar{\xi}_{\mathrm{H_{\gamma}}}(k) \right]},
\end{equation}
where $N_s$ is the number of simulations considered to compute these estimators. In particular, we take $N_s = 10000$ per hypothesis.

\subsection{The discriminator}
Following the description in \citet{Vielva2011}, we use the Fisher discriminant to distinguish between the two scenarios. This procedure is the optimal linear function of the measured quantities that maximizes the separation between the two probability distributions, $g(\tau|\mathrm{H_0})$ and $g(\tau|\mathrm{H_1})$, where $\tau$ is the value of the discriminant.

All required information to characterize both hypotheses is synthesized in the two vectors $\xi_{\mathrm{H_0}}$ and $\xi_{\mathrm{H_1}}$, so they are the estimators that we use as a starting point. The discriminator mechanism applied to $N$ signals corresponding, for instance, to the null hypothesis, leads to a set of $N$ numbers (called $\tau_{\mathrm{H_0}}$). They are the result of combining all the properties of $\mathrm{H_0}$ (i.e. $\xi_{\mathrm{H_0}}$, $\bar{\xi}_{\mathrm{H_0}}$ and $C_{\mathrm{H_0}}$), but taking into account the information related to $\mathrm{H_1}$ (i.e., $\bar{\xi}_{\mathrm{H_1}}$ and $C_{\mathrm{H_1}}$). Conversely, the Fisher discriminant applied to $N$ signals described by the alternative hypothesis provides a set of $N$ numbers (called $\tau_{\mathrm{H_1}}$) that are computed with the information of $\mathrm{H_1}$ but accounts for the overall properties of $\mathrm{H_0}$. 

In particular, we use the following expressions to calculate the $\tau_{\mathrm{H_{\gamma}}}$ values \citep[see e.g.][]{Barreiro2001, Martinez2002}:
\begin{equation} \begin{array}{ccc}
\tau_{\mathrm{H_0}} & = & \left( \bar{\xi}_{\mathrm{H_0}}-\bar{\xi}_{\mathrm{H_1}} \right)^t C_{\mathrm{tot}}^{-1} \xi_{\mathrm{H_0}} \\
\tau_{\mathrm{H_1}} & = & \left( \bar{\xi}_{\mathrm{H_0}}-\bar{\xi}_{\mathrm{H_1}} \right)^t C_{\mathrm{tot}}^{-1} \xi_{\mathrm{H_1}}, 
\end{array} \end{equation}
where $C_{\mathrm{tot}} = C_{\mathrm{H_0}} + C_{\mathrm{H_1}}$. 

In our case, $N=1000$ is the dimension of the sample considered for each hypothesis $\xi_{\mathrm{H_{\gamma}}}$, and the estimators $\bar{\xi}_{\mathrm{H_{\gamma}}}$ and $C_{\mathrm{H_{\gamma}}}$ are computed using $10000$ simulations, as we mentioned in the previous section.


\section{Forecast for data sets}
\label{sec:results}
In this section, we explore the scope of the methodology using simulations. We have used the 9-year best fit model of \textit{WMAP} to generate two sets of CMB simulations (that address both cases: extrema and random) following the steps described in Section \ref{sec:characterization}. As we are only interested in angular scales larger than $1^{\circ}$, the computation of $\mathrm{Q_r}$ maps and radial profiles has been performed at $N_\mathrm{side}=64$. 

As mentioned, two sets of $10000$ simulations have been employed to estimate the mean value of the vectors $\xi_{\mathrm{H_{\gamma}}}$, which contains the information of the $\mathrm{Q_r}$ radial profiles $\mu_{Q_\mathrm{r}}$, and the covariance matrices $C_{\mathrm{H_{\gamma}}}$. Other two sets of $1000$ simulations for each hypothesis have been considered in order to compute the distribution of the Fisher discriminants $\tau_{\gamma}$.

Let us recall some basic notions of statistical hypothesis testing. The significance level (or type I error), $\alpha$, is the probability of rejecting a given null hypothesis, $\mathrm{H_0}$, when $\mathrm{H_0}$ is true. The power of the test (or type II error) is the probability of rejecting the null hypothesis when the null hypothesis is false. Ideally, a good test would have a low significance level and high power. Furthermore, the p-value is the probability of obtaining an at least as extreme observation as the data when $\mathrm{H_0}$ is true. The null hypothesis is rejected if, and only if, the p-value obtained from the data is lower than the a-priori established significance level.

We have quantified the discrimination power between the two hypotheses by considering the significance level for a fixed power of the test of $(1-\beta)=0.5$, i.e., the fraction of the $\tau_{\mathrm{H_1}}$ values that are greater than the median value of the $\tau_{\mathrm{H_0}}$ distribution.

We have computed the significance level in the noise-free case considering different maximum angular distances $\theta_{max}$ for the profiles. We concluded that the significance level decreases drastically until an angular distance around $20^{\circ}$, where it reaches a value of approximately $1\%$ for larger angular scales.

\begin{figure*}
\includegraphics[scale=0.28]{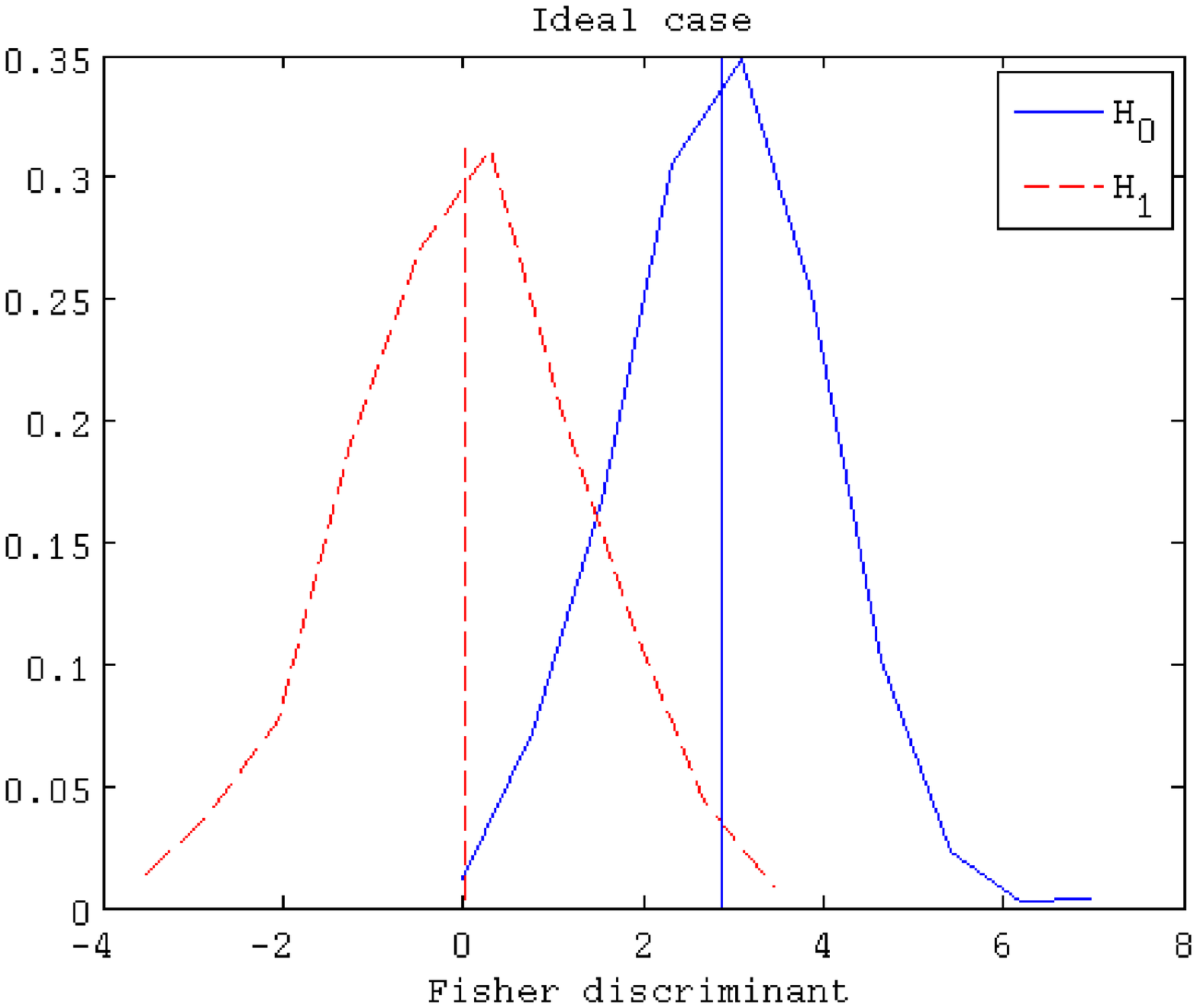}
\includegraphics[scale=0.28]{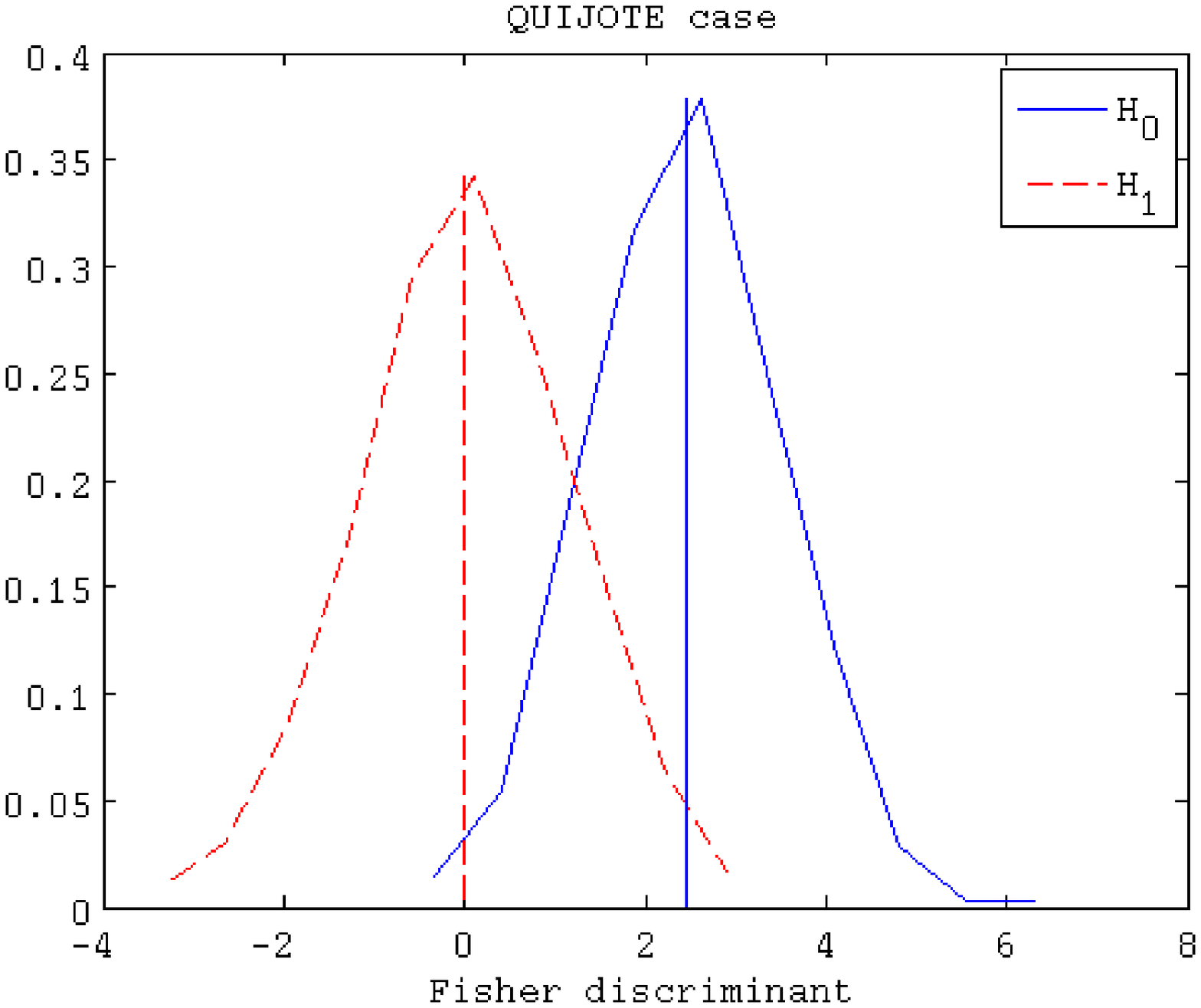}
\includegraphics[scale=0.28]{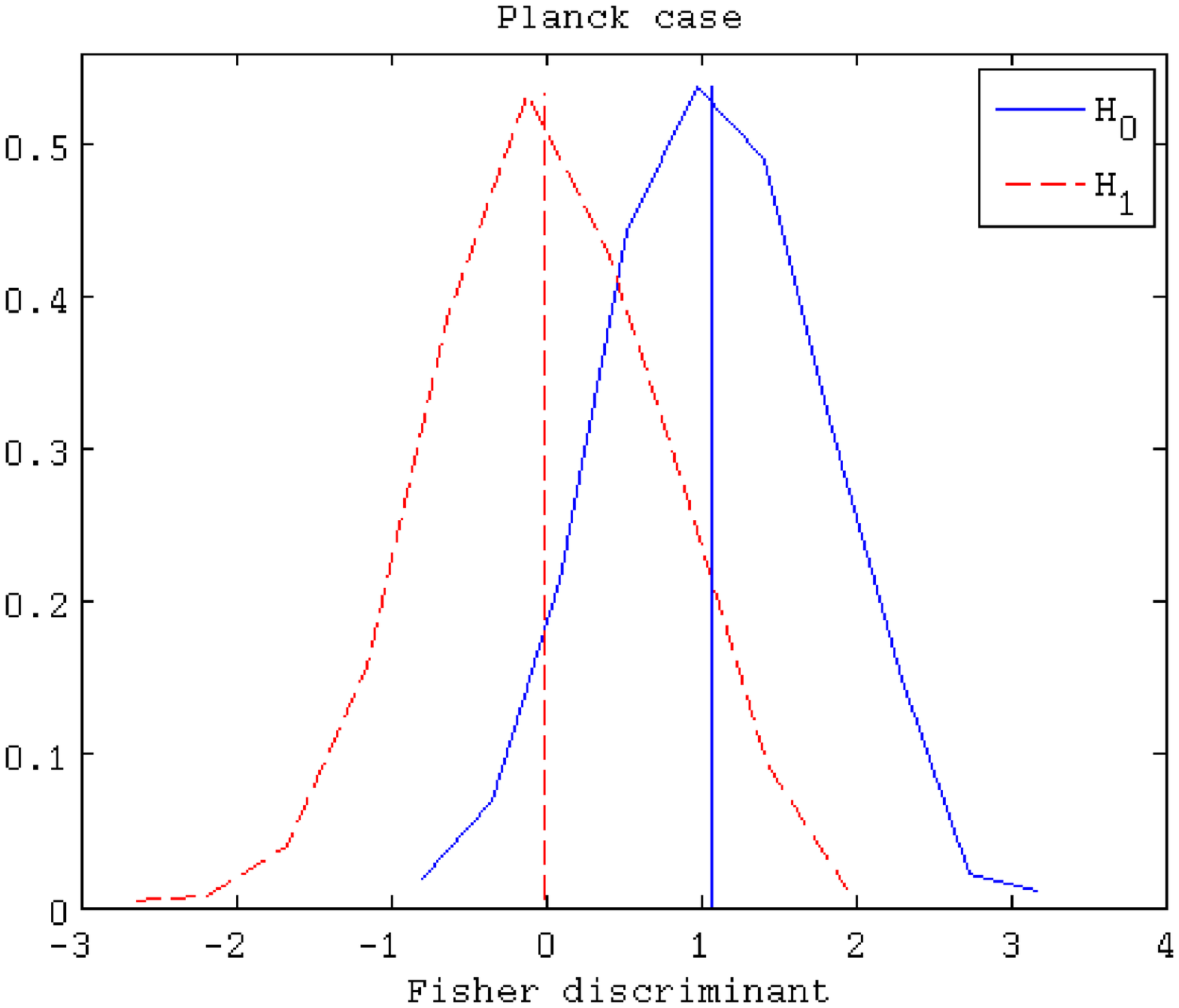}
\caption{Fisher discriminants for different level of white noise, from left to right: noise-free case, \textit{QUIJOTE}-like and \textit{Planck}-like levels. Solid blue lines correspond to the distribution of the Fisher discriminant for the null hypothesis ($\mathrm{H_0}$). The alternative hypothesis ($\mathrm{H_1}$) is represented by dotted red lines. The vertical lines mark the median values of each distribution. The significance levels at a power of the test of $0.5$ are: $0.010$, $0.015$ and $0.074$, respectively.}
\label{fig:discriminants}
\end{figure*}

In Figure \ref{fig:discriminants}, we show the distributions of the Fisher discriminants for three different cases. First of all, we present the noise-free case (noise amplitude $\sigma_{\mathrm{pol}} = 0$) in the left panel. Secondly, we plot a case with a noise level as expected in the \textit{QUIJOTE} experiment \citep{QUIJOTE2012} ($\sigma_{\mathrm{pol}} \approx 0.3\mathrm{\mu K}$ per pixel of $N_{\mathrm{side}} = 64$) in the middle panel. And, finally, we represent a third case considering the expected noise level in \textit{Planck} \citep{PlanckI2013} ($\sigma_{\mathrm{pol}} \approx 1\mathrm{\mu K} $ per pixel of $N_{\mathrm{side}} = 64$). The significance levels for a power of the test of $0.5$ are $0.010$, $0.015$ and $0.074$, respectively.

\begin{figure}
\includegraphics[scale=0.45]{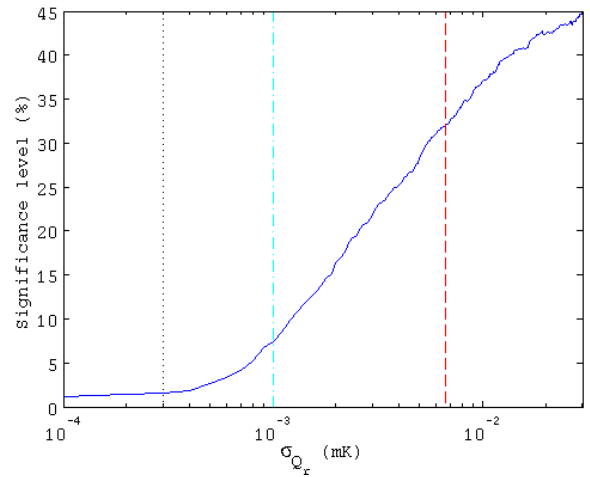}
\caption{Evolution of the significance level to reject the $\mathrm{H_1}$ hypothesis at a power of the test of $0.5$ with increasing instrumental-noise. The vertical lines, from left to right, correspond to the expected instrumental-noise levels of \textit{QUIJOTE}, \textit{Planck}, and the 9-year \textit{WMAP} data.}
\label{fig:significance}
\end{figure}

To provide a more general picture of the scope of the methodology, we have studied the evolution of the significance levels (for a power of the test of $0.5$) to discriminate between the two scenarios with increasing instrumental-noise level. The noise maps are computed as white-noise realizations. We show the result in Figure \ref{fig:significance}. The vertical lines, from left to right, correspond to the expected noise levels of \textit{QUIJOTE}, \textit{Planck} and the 9-year \textit{WMAP} data.  

\section{Application to the 9-year \textit{WMAP} data}
\label{sec:data}
In this section we show the results of applying the methodology to the case of 9-year \textit{WMAP} data. We have already mentioned that the CS is detected in the \textit{WMAP} temperature data with a SMHW filter at a wavelet scale of $R = 250$ arcmin and has an amplitude of $4.45\sigma$ in the wavelet coefficient map outside an extended version of the temperature mask. As in previous sections, we have considered a limit value of $4.45\sigma$ to select CMB realizations with a spot as extreme as the CS. However, the main difference between ideal simulations and those that we have used in this application is that the instrumental-noise pattern is not uniform. Therefore, we must ensure that the feature of our simulations has the same instrumental-noise pattern as that surrounded the CS position. 

Hence, we addopt a similar procedure to perform these simulations than in Section \ref{sec:characterization}. The only difference respect to the previous case is that, when we find a spot as extreme as the CS, we rotate the original ($\mathrm{T}$,$\mathrm{Q}$,$\mathrm{U}$) simulation in such a way that the feature is located at the CS coordinates ($l=209^{\circ}$,$b=57^{\circ}$). Furthermore, we take into account  the \textit{WMAP} beam window functions of the Q1, Q2, V1 and V2 differencing assemblies (DAs). Maps for different DAs are optimally combined into a single map using the $N_{obs}$ matrices supplied by the \textit{WMAP} team at the LAMBDA webpage\footnote{http://lambda.gsfc.nasa.gov/} \citep[see, for instance][]{Jarosik2011}.

For the treatment of the data we use a degraded version of the masks supplied by the \textit{WMAP} team in the 9-year release: the KQ75 and the polarization analysis one respectively. However, as the wavelet filter is applied over the masked temperature map, it is necessary to employ an extended version of this latter mask to exclude the contaminated regions before calculating the dispersion. 

\begin{figure}
\includegraphics[scale=0.45]{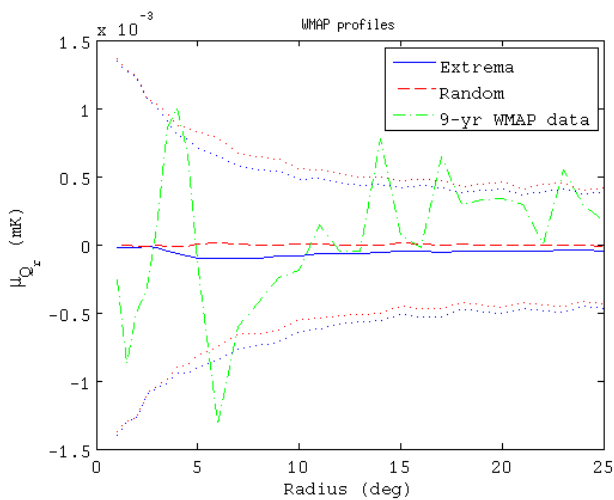}
\caption{Mean $\mathrm{Q_r}$ radial profile corresponding to \textit{WMAP}-like simulations. Result at extrema positions ($\mathrm{H_0}$), with an amplitude in the temperature maps, at least, as large as the one of the CS, are represented by the solid blue line. The profile computed at random positions ($\mathrm{H_1}$) in the temperature maps is plotted by a dashed red line. Their corresponding dispersion is shown by dotted lines. The dot-dashed green line corresponds to the Q+V \textit{WMAP} data profile.}
\label{fig:profile_WMAP}
\end{figure}

We show the $\mathrm{Q_r}$ mean radial profiles computed with $11000$ simulations in Figure \ref{fig:profile_WMAP}. The profile predicted by the null hypothesis ($\mathrm{H_0}$) is plotted by the solid blue line, whereas the dashed red line corresponds to the profile expected in $\mathrm{H_1}$. The dot-dashed green line represents the Q+V \textit{WMAP} data profile. 

\begin{figure}
\includegraphics[scale=0.45]{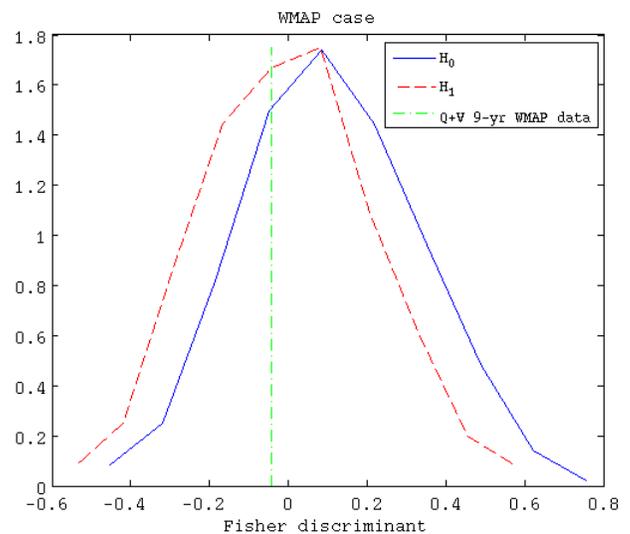}
\caption{Fisher discriminant for the \textit{WMAP} case. The solid blue line corresponds to the distribution of the Fisher discriminant for the null hypothesis ($\mathrm{H_0}$). The alternative hypothesis ($\mathrm{H_1}$) is represented by dashed red lines. The significance level at a power of the test of $0.5$ is $0.32$. The vertical dot-dashed green line marks the discriminant value associated to the Q+V \textit{WMAP} data, $\tau_{\mathrm{data}} = -0.04$.}
\label{fig:discriminant_WMAP}
\end{figure}

The results of applying the Fisher discriminant methodology are shown in Figure \ref{fig:discriminant_WMAP}. The distribution of the Fisher discriminant for the null hypothesis ($\mathrm{H_0}$) is represented by a solid blue line. The dashed red line corresponds to the distribution of the alternative hypothesis $\mathrm{H_1}$. The significance level (at a power of the test of $0.5$) is $0.32$, which indicates that the hypothesis test is really bad in this case. The instrumental-noise level is too high to obtain a strong finding. In fact, given that significant level, we are unable to reject the null hypothesis independently of the obtained p-value (since the probability of rejecting $\mathrm{H_0}$, being true, is $32\%$). The vertical dot-dashed green line marks the discriminant value associated to the Q+V \textit{WMAP} data, $\tau_{\mathrm{data}} = -0.04$, so that we obtain a fraction of $0.26$ of simulations with a discriminant value as extreme as $\tau_{\mathrm{data}}$.

\section{Conclusions}
\label{sec:Conclusions}

We present a procedure in terms of hypothesis testing to distinguish between two different scenarios that might be behind the nature of the CS. On the one hand, the case in which this feature is merely a very extreme fluctuation of the Gaussian isotropic random field compatible with the standard inflationary model predictions. On the other hand, the proposal which considers the CS as due to a contribution that does not present a correlated pattern between temperature and polarization (such as topological defects), and is superimposed to the standard Gaussian field. The basis of the method consists in optimizing the differences in the cross-correlation patterns between the temperature and the polarization E-mode, estimated via the $\mathrm{Q_r}$ Stokes parameter.

We have explored the possibilities of this methodology in terms of the instrumental noise levels. For experimental sensitivities that can be reached at present, we have obtained promisingly low significance levels (at a power of the test of $0.5$) to reject the alternative hypothesis. In particular, the estimation of this value is $0.010$, $0.015$ and $0.074$ for an ideal noise-free experiment, \textit{Planck}-like and \textit{QUIJOTE}-like noise levels, respectively. These results are very similar to those obtained by \citet{Vielva2011} in the case where full-sky coverage is assumed, so we are not losing effectiveness due to the considering of an exclusion mask.

Furthermore, we have applied the method to the particular case of \textit{WMAP} data, obtaining a significant level of $0.32$. The instrumental noise level is too high to discriminate between the two hypotheses. The estimated significance levels have been computed assuming that the temperature is anomalous, but, in this case, the analysis of the polarization data does not add anything else with respect to the result obtained by considering only temperature data. However, it is expected that this method will be useful when dealing with data sets provided by the current generation of CMB polarization experiments.


\section*{acknowledgments}
The authors thank Rita Bel\'en Barreiro and Airam Marcos-Caballero for comments and useful discussions. We acknowledge partial financial support from the Spanish \textit{Ministerio de Econom\'ia y Competitividad} Projects AYA2010-21766-C03-01, AYA2012-39475-C02-01 and Consolider-Ingenio 2010 CSD2010-00064. RFC thanks financial support from Spanish CSIC for a \textit{JAE-predoc} fellowship, co-financed by the European Social Fund. The authors acknowledge the computer resources, technical expertise and assistance provided by the \textit{Spanish Supercomputing Network} (RES) node at Universidad de Cantabria. We acknowledge the use of \textit{Legacy Archive for Microwave Background Data Analysis} (LAMBDA). The HEALPix package was used throughout the data analysis \citep{Gorski2005}.

\bibliographystyle{mn2e}
\bibliography{citas_stacking}

\end{document}